\documentclass[]{revtex4-1}
\usepackage{amsmath}    
\usepackage{epstopdf}
\usepackage{graphicx}
\usepackage{verbatim}   
\usepackage{color}      
\usepackage{hyperref}   
\usepackage{lineno} 
\usepackage{bm} 
\usepackage{multirow}
\usepackage{csquotes}
\usepackage{appendix}
\usepackage{caption}
\usepackage{subcaption}
\usepackage{nomencl}
\makenomenclature
\newcommand{\classoption}[1]{\texttt{#1}}

\newcommand{\pt}{\ensuremath{p_{\mathrm{T}}}}

\expandafter\ifx\csname package@font\endcsname\relax\else
\expandafter\expandafter
\expandafter\usepackage
\expandafter\expandafter
\expandafter{\csname package@font\endcsname}%
\fi
\DeclareRobustCommand\substyle{\name@idx{document substyle}}%
\DeclareRobustCommand\classoption{\name@idx{document class option}}%
\DeclareRobustCommand\classname{\name@idx{document class}}%
\def\name@idx#1#2{
	{\ttfamily#2}%
	\index{#2\space#1=\string\ttt{#2}\space#1}\index{#1>#2=\string\ttt{#2}}%
}

\newcommand{\ts }{transverse spherocity }
\newcommand{\pts }{$p_T$ }
\newcommand{\sqt }{$\sqrt{s_{NN}}$ = 5.44 TeV }
\newcommand{\pio }{$\pi^{\pm}$ }
\newcommand{\kao }{$K^{\pm}$ }
\newcommand{\pro }{$p^{\pm}$ }
\newcommand{\cent }{centrality }
\newcommand{\so }{$S_o$ }

\begin{document}

	\pagenumbering{arabic}
	\title{Centrality and Transverse Spherocity dependent study of charged-particle production in Xe-Xe collisions at $\sqrt{s_{NN}}$ = 5.44 TeV using PYTHIA8 Angantyr and AMPT models }
\author{ Randhir Singh$^{1}$\footnote{E-mail: randhir.singh@cern.ch},  Renu Bala$^{1}$\footnote{E-mail: Renu.Bala@cern.ch}, Sanjeev Singh Sambyal$^{1}$\footnote{E-mail: sanjeev.singh.sambyal@cern.ch}}

\affiliation{Department Of Physics, University of Jammu, Jammu, India$^{1}$}

	\begin{abstract}
Transverse Spherocity is an event structure variable which provide an effective way to disentangle the data into hard and soft components of the processes corresponding to events with small and large numbers of multi-parton interactions (MPI), respectively. Recent experimental results in small systems from the LHC suggest the importacnce of transverse spherocity variable in the classification of the events. In this contribution, we have studied the dynamics of identified particle production in Xe-Xe collisions at $\sqrt{s_{NN}}$ = 5.44 TeV using A Multi-Phase Transport Model (AMPT) and the recently developed Angantyr model, which is incorporated within  PYTHIA8. A study of the transverse momentum spectra of the dentified particles are  presented for soft (isotropic) and hard (jetty-like) events in different centrality intervals.
 
	\end{abstract}
	\maketitle

\section{Introduction}
\label{section1}
The study of the behaviour of the matter at extreme conditions of temperature and energy density is important to understand the processes involved in the phase transition of the matter in Quantum Chromodynamics (QCD). Experiments like Relativistic Heavy Ion Collider (RHIC) at BNL, USA and Large Hadron Collider at CERN, Geneva, Switzerland are specially designed to study the matter at these extreme conditions. In these experiments, the results have suggested that a strongly interacting de-confined state of quarks and gluons, also called Quark-Gluon plasma (QGP), is formed in ultra-relativistic heavy-ion collisions. This plasma is a very hot and dense state and exist for a very short duration of time. The relevant degrees of freedom in the QGP phase are quarks and gluons instead of mesons and baryons (which are color neutral confined state). After the formation this QGP plasma cools and expands hydrodynamically and reaches first to the chemical freeze-out (the hadron abundances are fixed) phase and then reaches to kinetic freeze-out (the hadron momenta are fixed) phase.\\
\\
The observable which is important to cahracterize the properties of the created matter in these collisions is the multiplicity of the charged particles. This is due to the fact that the partcle production is effected by the initial energy density reached in the collision. The colliding nuclei are extended objects. Therefore, in a collision, the region of overlap between the two colliding nuclei varies from collision to collision. This degree of overlap is expressed by the term impact parameter (denoted as "b"). The impact parameter cannot be measured directly. To overcome the difficulty, a different parameter called centrality is used. The centrality characterises the initial overlapping region geometry, which corresponds to the number of participating nucleons N$_{part}$ and binary collisions N$_{coll}$. N$_{part}$ and N$_{coll}$  will be different for similar overlap in collisions of nuclei of different sizes.\\
\\
Transverse spherocity is an event shape variable which has been introduced recently. The recent development in event shape variables provides an alternate way to characterise the high multiplicity events in small systems considering particle production, event multiplicity, and event shape variable together. Event shape observables provide information about the energy distribution in an event. The event shape observables, calculated event-by-event allows one to isolate jetty-like (high p$_T$ jets) and isotropic (partonic scattering with low Q$^2$) events \cite{Morsch}. The ALICE, CMS, and ATLAS experiments have done the studies based on transverse spherocity \cite{Chatrchyan,Abelev}. These studies involved the understanding of event shape as a function of charged particle multiplicity of the event. The results from these studies show that for
high multiplicity, the events were more isotropic. After studying extensively in small systems \cite{ALICE:2019dfi}, the use of transverse spherocity parameter in heavy-ion collisions systems may reveal new and unique results from heavy-ion collisions where the production of a QGP is well established fact.\\
\\
Transverse momentum (p$_T$) spectra of the charged particles produced in the heavy-ion collisions carry essential information about the thermal
nature of the interacting system \cite{Cleymans:2012ya,Cleymans:2011in,Mallick:2020ium, Mallick:2021wop, Mallick:2021hcs, Prasad:2021bdq, Prasad:2022zbr, Rath:2018ytr}. The Maxwell-Boltzmann distribution law tells us that the p$_T$ spectra are related to the temperature of the system formed in these collisions. At low to intermediate p$_T$ (up to 10 GeV/$c$), the collective expansion of the system governs the charged particle production, which is observed in the shapes of single-particle transverse-momentum spectra \cite{ALICE:2015dtd,ALICE:2018vuu} and multi-particle correlations \cite{Heinz:2013th}. At high p$_T$ (typically above 10 GeV/$c$), parton fragmentation governs the particle production. This production is effected by the amount of energy loss that the partons suffer when propagating in the medium. Moreover, the number of charged particles (N$_{ch}$) produced in these ultra-relativistic collisions are also related to the temperature and energy density of the system created. The ratios of yields of identified hadrons are also important to understand to the processes involved in the production of hadron. The ratio of proton to pion (p/$\pi$) and kaon to pion (K/$\pi$) characterize the relative baryon and meson production, respectively.\\
\\
In this work, we have explored all the above-mentioned acpects related to particle production. We have also used transverse spherocity for the first time in Xe-Xe collisions at $\sqrt{s_{NN}}$ = 5.44 TeV using A Multi-Phase Transport Model (AMPT)\cite{Lin:2004en} and Angantyr model which is incorporated in PYTHIA8  . The purpose of the present analysis is to study the dynamics of heavy-ion collisions using transverse spherocity and collision centrality in Xe-Xe collisions. Thus, we have used the PYTHIA8 and the string melting parameterization available in AMPT model. We hope that this would drive experimentalists to pursue such study in experiments at RHIC and the LHC. The contribution is arranged as follows. We begin with a brief introduction and motivation for the study in Section-\ref{section1}. In Section-\ref{section2}, the detailed analysis methodology along with brief description about AMPT and PYTHIA (Angantyr) are given. Section-\ref{section3}  discusses about the results for the charged-particle pseudorapidity density, \pts spectra of \pio, \kao and \pro in different \cent and \ts ranges. Finally the results are summarized in Section-\ref{section4}. 

\section{EVENT GENERATION AND ANALYSIS	METHODOLOGY}
\label{section2}
This section will be devoted to discussions related to the event generators, AMPT and PYTHIA8 (Angantyr). Around 0.5 Million Xe-Xe events have been generated for both event generators at  $\sqrt{s_{NN}}$ = 5.44 TeV. A discussion will also be presented for transverse spherocity.\\

\subsection{PYTHIA8 (Angantyr)}

PYTHIA8 is general purpose Monte Carlo event generator that has been quite successful in the study of elementary particle collisions. It has been extensively used to simulate proton-proton, proton-lepton collisions to understand the dynamics of strong and electroweak processes extending from high momemtum transfer regions (perturbative scales) to the scales around $\Lambda_{QCD}$ (Lattice QCD scales). Recently efforts have been made to modify PYTHIA8 that made it possible to simulate ultra-relativistic heavy-ion collisins like, proton-nuclei (pA) and nuclei-nuclei (AA). PYTHIA8 incorporates what is called "Angantyr" which basically superpose many pp collisions to make a single heavy-ion collision. This way a bridge is formed between heavy ion and high energy hadron phenomenology. An important point is to note that the assumption of the formation of a hot thermalised medium is not included in the Angantyr as the production mechanisms is same as in small collision systems. Thus it can be used to differentiate the effects of collective and non-collective behaviour in heavy-ion collisions.\\
\\
The Angantyr model is basically inspired by the old Fritiof model and the notion of “wounded” nucleons, but with the inclusion of effects of hard partonic interactions. To calculate the number of wounded nucleons the use of Glauber model is made, with the consideration of fluctuations in the nucleon-nucleon(NN) interactions to differentiate non-diffractively and diffractively wounded nucleons. In the model, the fluctuations of neuclonic wavefunction are accommodated as fluctuations in the nuclei radius. Nucleons  inside nuclei are distributed randomly according to a Glauber formalism. Every nucleon is then identified as either wounded or spectator. The interections between wounded nucleons in projectile and target are classified as elastic, ND, secondary ND, SD and double-diffrative(DD) which basically depends upon the interaction probability. All the sub-events generated at the parton level are stacked together to represent one pA or AA collision event. 

In PYTHIA8 hadronization is done via the Lund string fragmentation model. In this model, the Lund area law \cite{Andersson:2001yu} provides the probability of creating hadrons from an initial state of partons. The produced partons are connected to the beam remnants through color fluxtubes or strings which stores potential energy. As the partons move away from each other, the string breaks causing the formation of new quark-antiquark pair. This process continues until the string pieces reduces to very small pieces, which are
recognized as shell hadrons. In this scheme, reconnection of strings between the partons happens in such a way that strings length decreases; resulting to the decrease in the particle production and hence multiplicity.\\

\subsection{A Multi-Phase Transport (AMPT) model}

The AMPT model has four main components : the initial conditions, partonic interactions, the conversion from the partonic to the hadronic matter, and hadronic interactions. In case of default AMPT model, initial conditions for heavy ion collisions at RHIC are carried from the HIJING model \cite{Wang:1990qp,Wang:1991hta,Gyulassy:1994ew}. The production of particles are described either as hard and soft components. The hard components are the ones that involve a momentum transfer larger than a cutoff momentum, $p_0$, value while soft components are the ones having momentum transfer below the cutoff value. The hard component is evaluated in the perturbative Quantum Chromodynamics domain (pQCD) using parton distribution functions in a nucleus. These processes contribute to the production of minijets partons. The soft component is discussed in non-perturbative region and are modeled by the formation of strings. The excited strings then decay independently following Lund JETSET fragmentation model. The energy density in the default AMPT model can be very high in heavy ion collisions. To incorporate this effect, the AMPT model is extended to include the string melting mechanism. The interactions between partons are described using equations of motion which can be approximately written as the Boltzmann equations. The Boltzmann equations are then solved by using Zhang’s parton cascade (ZPC) \cite{Zhang:1997ej}, in which two partons undergo
scattering whenever they are within a minimum closest distance. In the String Melting version of AMPT (AMPT-SM), colored strings are melted to form low momentum partons which take place at the start of the ZPC. Corresponding to two different initial conditions, hadronization is also done via two different mechanisms. In case of default AMPT model, the
minijets and their remaining parent nucleons coexist and after partonic interactions they together form new excited strings. Lund string fragmentation model describes the hadronization of these strings. In case of the AMPT model with string melting, the conversion of strings into soft partons takesplace, and after that hadronization happens via simple quark coalescence model. In the coalescence model, hadronization is done by combining three closest quarks (antiquarks) into a baryon (antibaryon) and two closest partons into a meson. The hadrons thus produced undergo final evolution via meson-meson, meson-baryon and baryon-baryon interactions described by the relativistic transport mechanism. The quark coalescence mechanism for hadronization well explains the spectra at the mid-p$_T$ regions and particle flow \cite{Greco:2003mm,Fries:2003vb,Fries:2003kq}. Thus, for our work, we have used AMPT-SM mode (AMPT version 2.26t7) with the default settings.\\
\\
\begin{figure}[h!]
	\centering
	\includegraphics[scale=0.35]{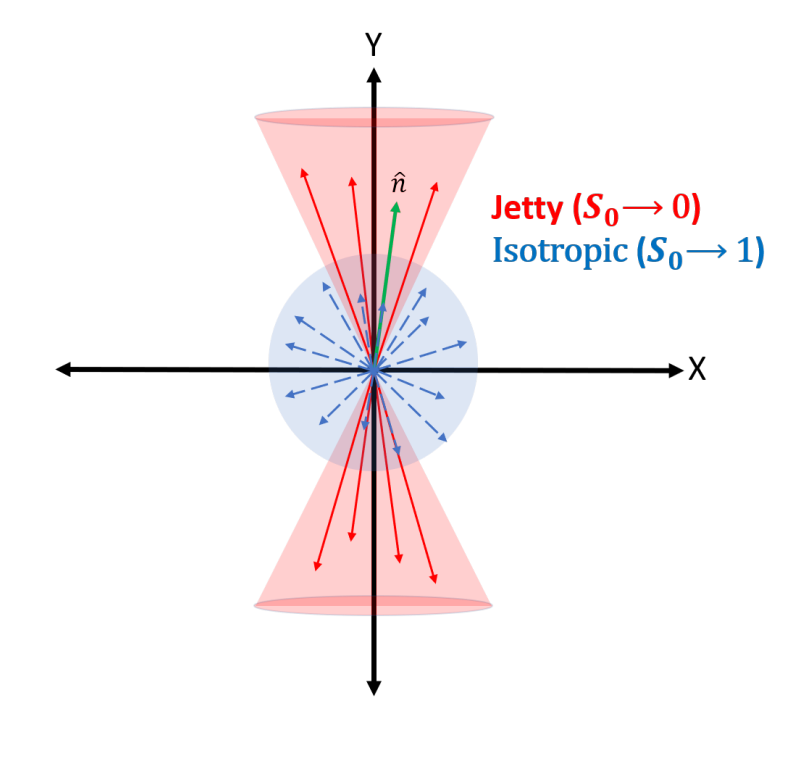}
	\caption{Jetty and isotropic events in the transverse plane\cite{Khuntia:2018qox}. }
	\label{spherocity1}
\end{figure}
\subsection{Transverse Spherocity}
Transverse spherocity $(S_o)$ is the property of an event and is defined using a unit vector $\hat{n}(n_T,0)$ which minimizes the ratio \cite{Cuautle:2014yda,Ortiz:2015ttf} and chosen from all possible unit transverse vectors:

\begin{equation}
\centering
S_{o} = \frac{\pi^{2}}{4}  \underset{\widehat{n}=(n_{x},n_{y},n_{y})}{min}\Bigg(\frac{\sum_i\vert\vec{p}_{T_i} \times \widehat{n}_{T}\vert}{\sum_{i} p_{T_i}}\Bigg)^{2}
\end{equation}

The factor, $\pi^{2}/4$ normalises the minimised value to 1 for isotropic case. As a result, the value of transverse spherocity run from 0 to 1, as the distribution of particles deviates from the jetty-like to an isotropic structure respectively, i.e.

\begin{equation}
\centering
S_{o} = \begin{cases}
0 & \text{“pencil-like” limit (hard events)}\\
1 & \text{“isotropic” limit (soft events)}\\
\end{cases}
\end{equation}
Fig.\ref{spherocity1} shows the distribution of particles in both the scenarios. 

\begin{figure}[h!]
	\centering
	\includegraphics[scale=0.50]{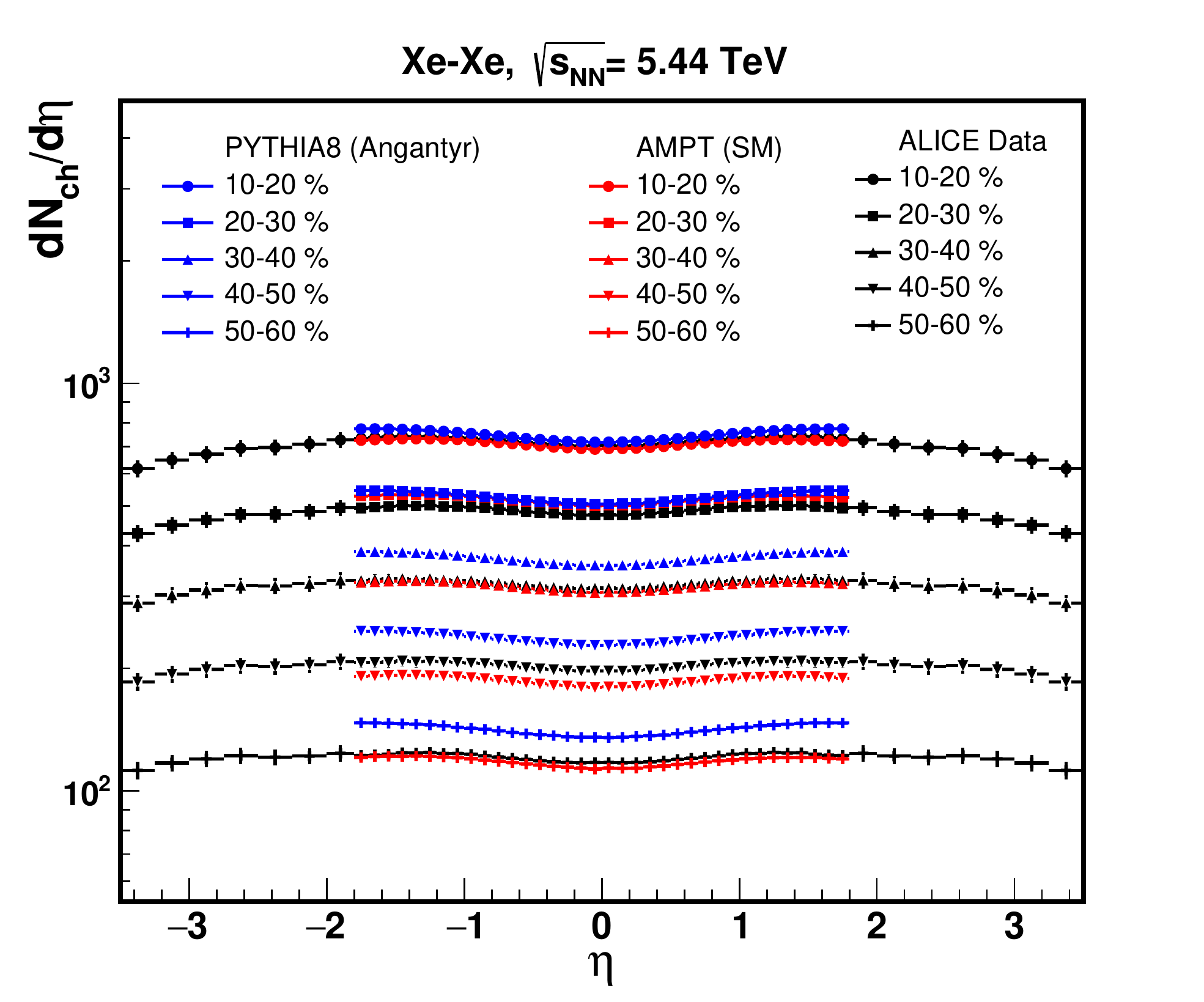}
	\caption{Charged-particle pseudorapidity density for different centrality classes over a broad $\eta$ range in \textbf{Xe-Xe} collisions at \sqt calculated with AMPT and PYTHIA8. The results also compared with the results from the ALICE Experiment at the LHC at CERN. }
	\label{dNchdEta}
\end{figure}
\begin{table}[h!]
	\centering
	\resizebox{12cm}{!}{\begin{tabular}{|l|l|l|l|l|}
		\hline
		\multirow{2}{*}{\textbf{Centrality }} &
		\multicolumn{2}{c|}{\textbf{AMPT}} & \multicolumn{2}{c|}{\textbf{PYTHIA8}}\\
	     \cline{2-5}
		& $b_{min}$ & $b_{max}$ & $E_T^{min}$& $E_T^{max}$  \\
		\hline
		0-10\% & 0 & 4.507  & 1031.47 & 2000  \\
		\hline
		10-20\% &4.507 & 6.37525 & 729.13 & 1031.47  \\
		\hline
		20-30\% & 6.37525 & 7.80479 & 502.604& 729.13\\
		\hline
		30-40\% & 7.80479 & 9.00297 & 330.139 & 502.604\\
		\hline
		40-50\% & 9.00297 & 10.0606 & 202.442& 330.139\\
		\hline
		50-60\% & 10.0606 & 11.0097 & 111.821& 202.442\\
		\hline
		60-70\% & 11.0097 & 11.8943 & 53.8031&111.821\\
    	\hline
    	
\end{tabular}}
	\caption{Table showing the values of impact parameter for AMPT and summed transverse energy for PYTHIA8 for different centrality bins.}
	\label{centrality}
\end{table}

\section{RESULTS}
\label{section3}
\begin{figure}[h!] 
	\centering
	\begin{subfigure}[b]{0.45\linewidth}
		\includegraphics[scale=0.20]{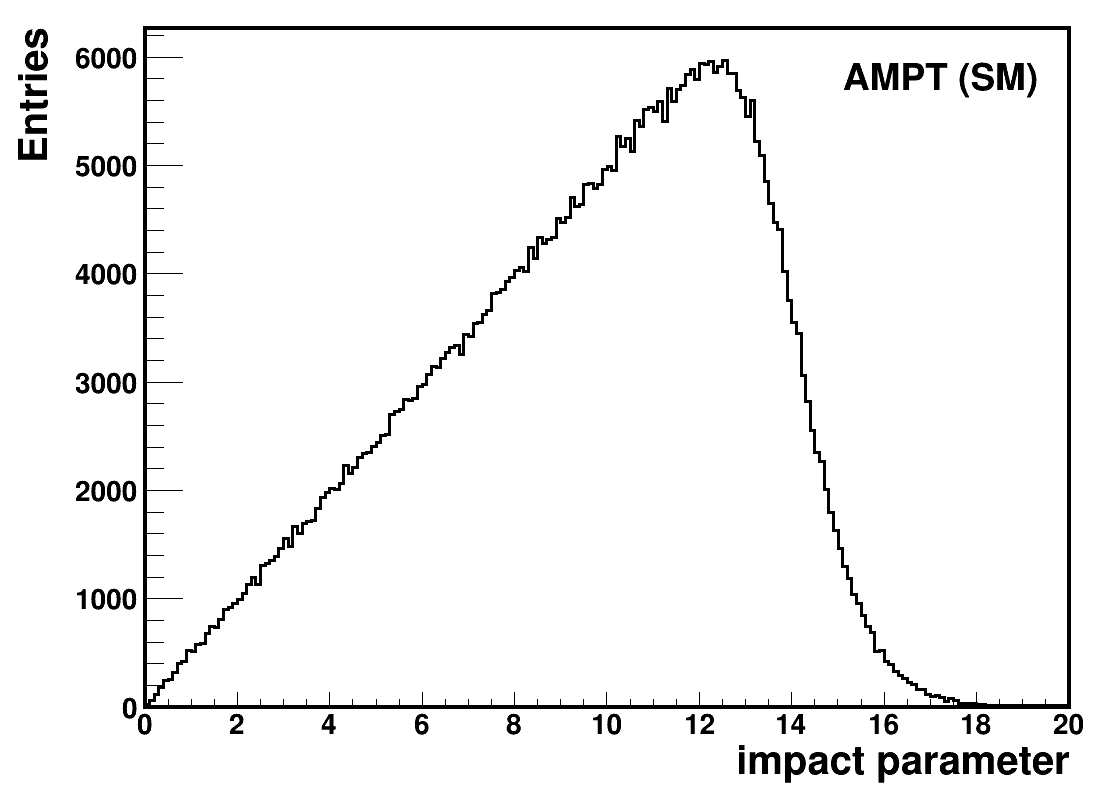}
		\caption{AMPT: String Melting.}
		\label{distributionsa}	
	\end{subfigure}
	\begin{subfigure}[b]{0.45\linewidth}
		\includegraphics[scale=0.20]{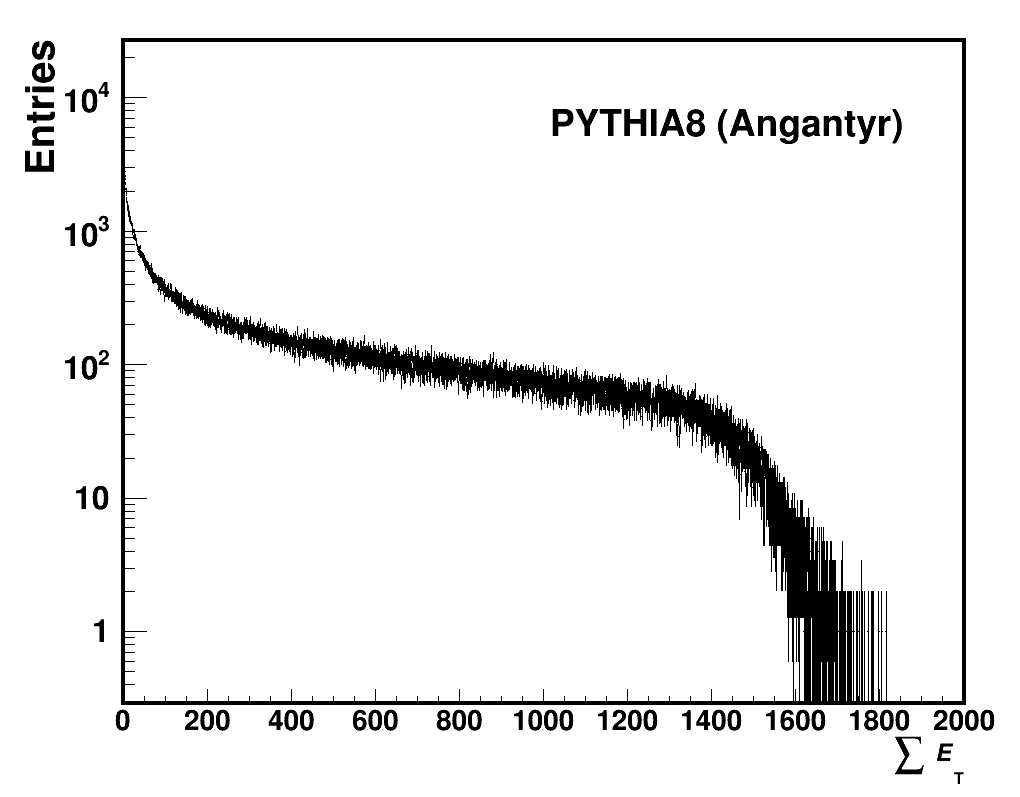}
		\caption{PYTHIA8 Angantyr}
		\label{distributionsb}
	\end{subfigure}
	\caption{(a) Event distribution as a function of impact parameter for AMPT and (b) Event distribution as a function of Summed Transverse energy ($\sum E_T$) for PYTHIA8 Angantyr model.}
	\label{distributions}
\end{figure}

\begin{figure}[h!] 
	\centering
	\begin{subfigure}[b]{0.45\linewidth}
		\includegraphics[scale=0.42]{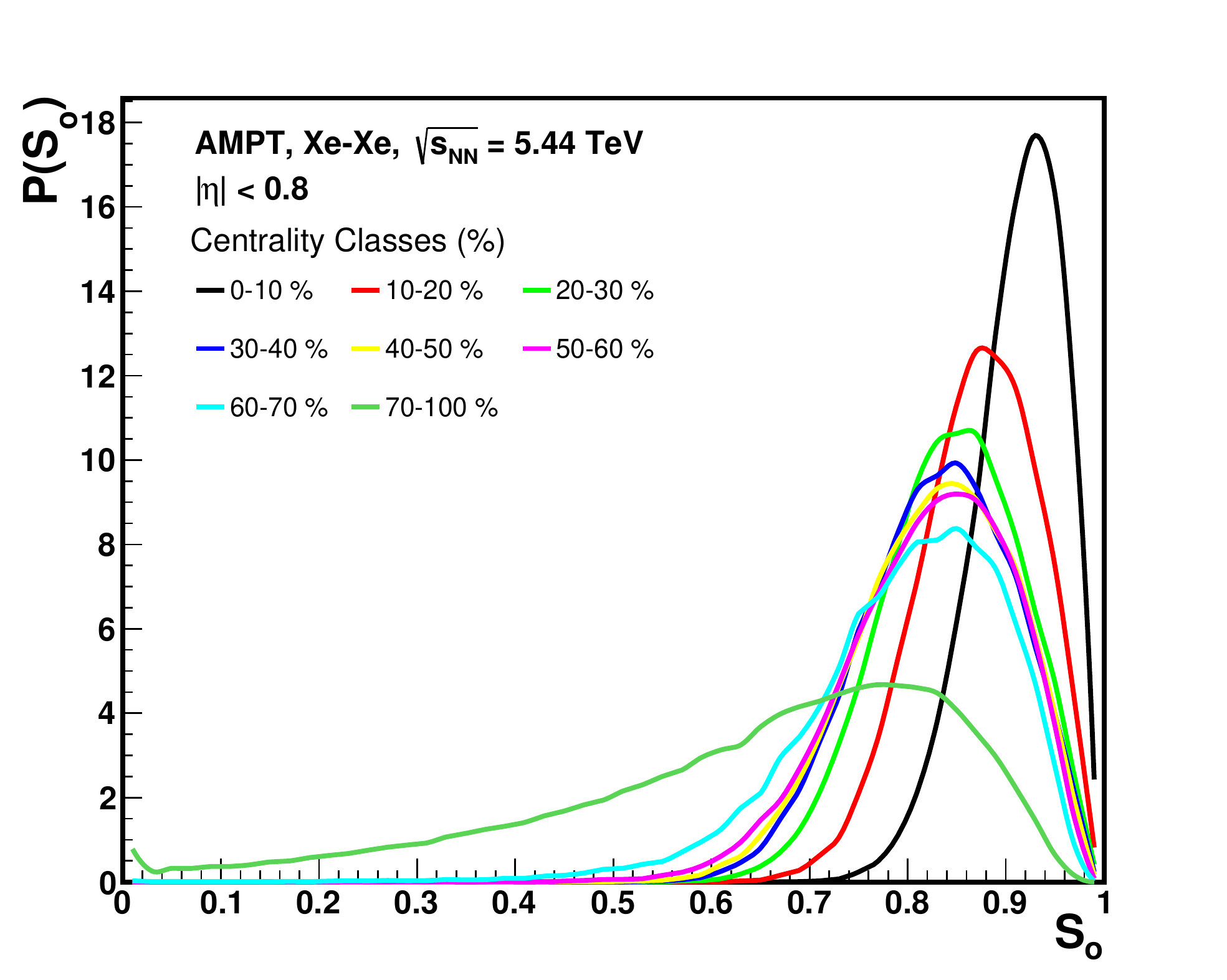}
		\caption{AMPT: String Melting.}
		\label{AMPTdistributions}	
	\end{subfigure}
	\begin{subfigure}[b]{0.45\linewidth}
		\includegraphics[scale=0.42]{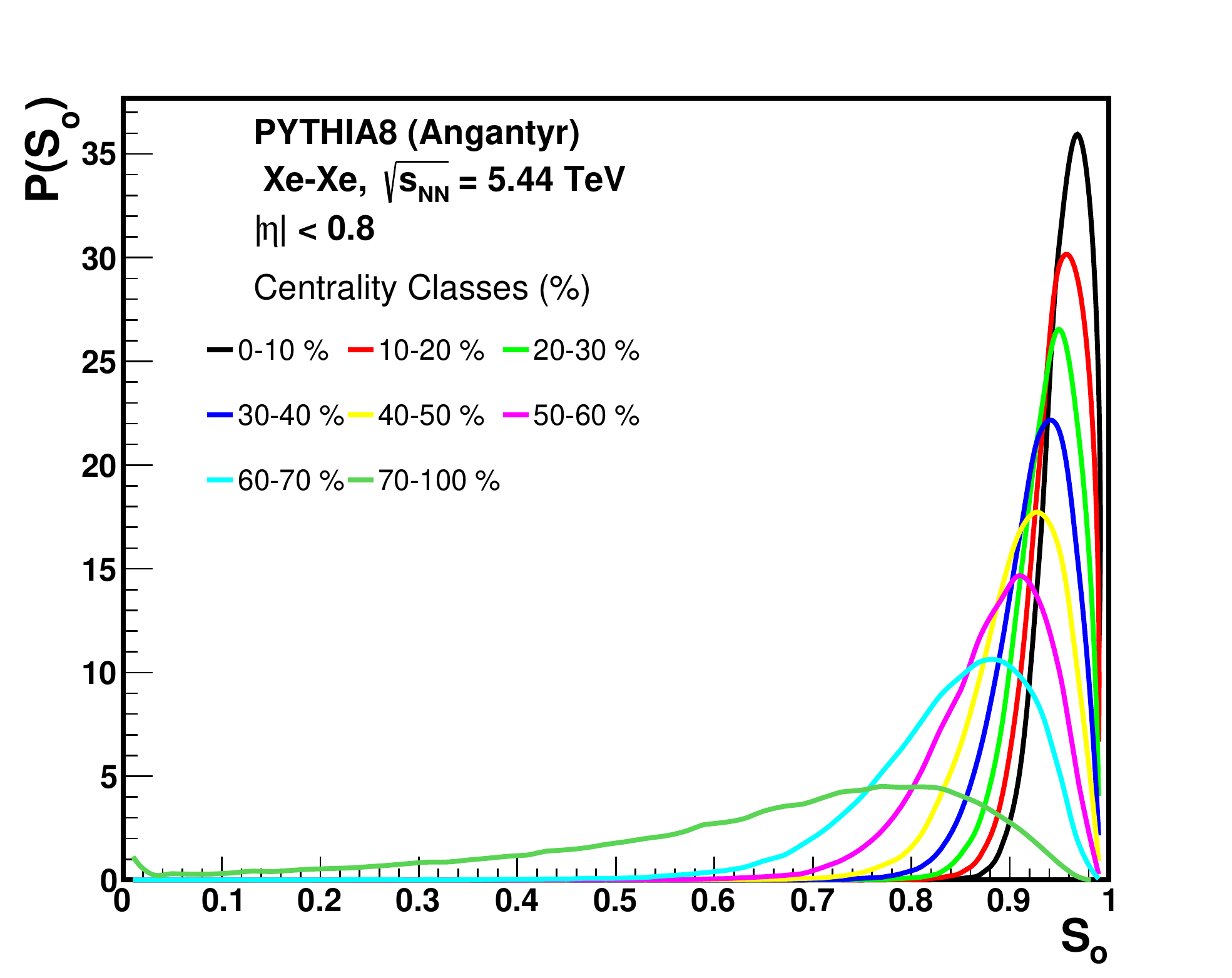}
		\caption{PYTHIA8 Angantyr}
		\label{Pythiadistributions}
	\end{subfigure}
	\caption{(a) Event distribution as a function of impact parameter for AMPT and (b) Event distribution as a function of Summed Transverse energy ($\sum E_T$) for PYTHIA8 Angantyr model.}
	\label{spherocity_distributions}
\end{figure}

In figure \ref{dNchdEta}, the charged-particle multiplicity density $dN_{ch}/d\eta$ as a function of pseudorapidity for different centrality classes are presented for AMPT and PYTHIA8. The centrality intervals for the AMPT are defined using the impact parameter values as can be seen in table \ref{centrality} whereas for PYTHIA8 we define the centrality intervals which is based on the summed transverse energy ($\sum E_T$) in the pseudorapidity interval [-0.8, 0.8]. Figure \ref{distributionsa} shows the distribution of events as a function of impact parameter for AMPT generator while Figure \ref{distributionsb} shows event distribution for PYTHIA8 event generator as a function of summed transverse energy ($\sum E_T$) \cite{Bierlich:2018xfw}. The $dN_{ch}/d\eta$ values are also compared with the results from the experimental study with the ALICE Detector at the LHC at CERN \cite{ALICE:2018cpu}. From the figure, it can be seen that the AMPT describe the data fairly well at low centralities, and a slight overestimate at high centralities. On the other hand, PYTHIA8 produces the shape pretty well but slightly overestimates the data in the small centrality bins but this overestimation increases in higher centrality bins. For Angantyr, we find that the fluctuations as well as the distinction between primary and secondary absorption-damaged nucleons have a fairly significant impact on the final state multiplicity. It is to be noted that AMPT presumes that a hot dense medium is formed in which a collective expansion of a thermalised medium occurs whereas PYTHIA lacks a mechanism to reproduce the collective effects seen in pp collisions, and therefore the Angantyr model lacks the assumption of production of a medium as well.
The \pts spectra of particles has been studied in jetty and isotropic limits of transverse spherocity intervals for centrality classes. In Figure \ref{spherocity_distributions} the \ts distribution for AMPT \ref{AMPTdistributions} and PYTHIA8 \ref{Pythiadistributions} generators in different centrality classes for Xe-Xe collisions at \sqt are presented. From the figure, it is observed that small centrality (high-multiplicity) events are more towards isotropic in nature whereas high centrality (low multiplicity) events are towards the jetty side. The means that the peak of the \ts  distribution shifts towards jetty events with centrality, which shows that more central events contribute to the softer events and vice-versa. The process of isotropization in a many-particle final state occurs through multiple interactions between the quanta of the system. If the final state multiplicity in an event is higher, the probability that the event will become isotropic is also higher. Therefore, the differential
study of particle production as a function of centrality and \ts classes has great importance to understand the particle production mechanism. As \ts distribution depends on centrality and ultimately to the charged-particle multiplicity, therefore to define the jetty and isotropic limits of \ts, the 20$\%$ events of the extremum of the spherocity distribution are considered. The cuts for jetty and isotropic events vary for different centrality classes are shown in Table.\ref{values}.\\

\begin{figure}[h!] 
	\centering
	\begin{subfigure}[b]{0.30\linewidth}
		\includegraphics[width=6cm,height=5cm]{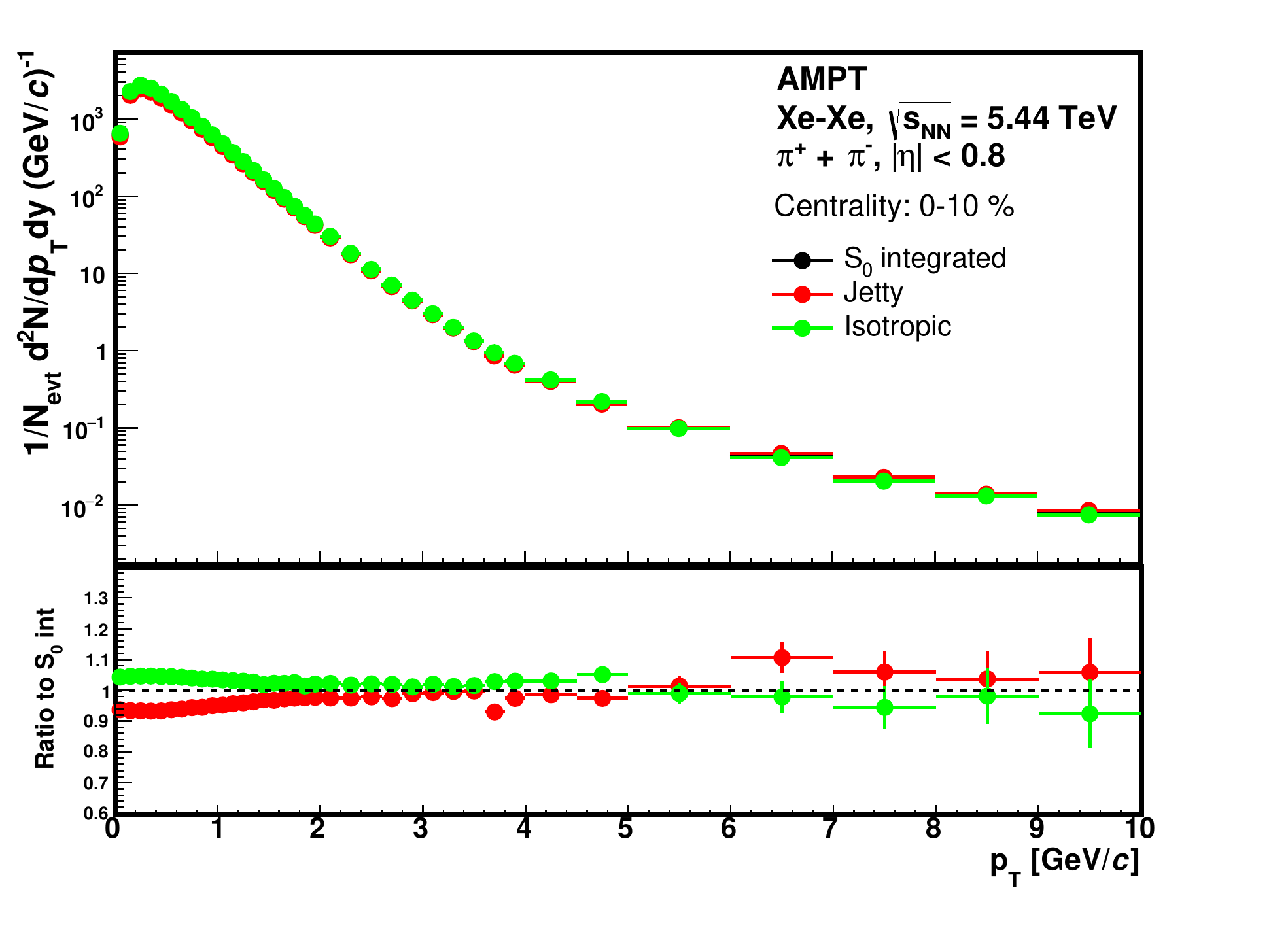}
		\caption{\textbf{AMPT: \pio}}
		\label{AMPT_pion0a}	
	\end{subfigure}
	\begin{subfigure}[b]{0.30\linewidth}
		\includegraphics[width=6cm,height=5cm]{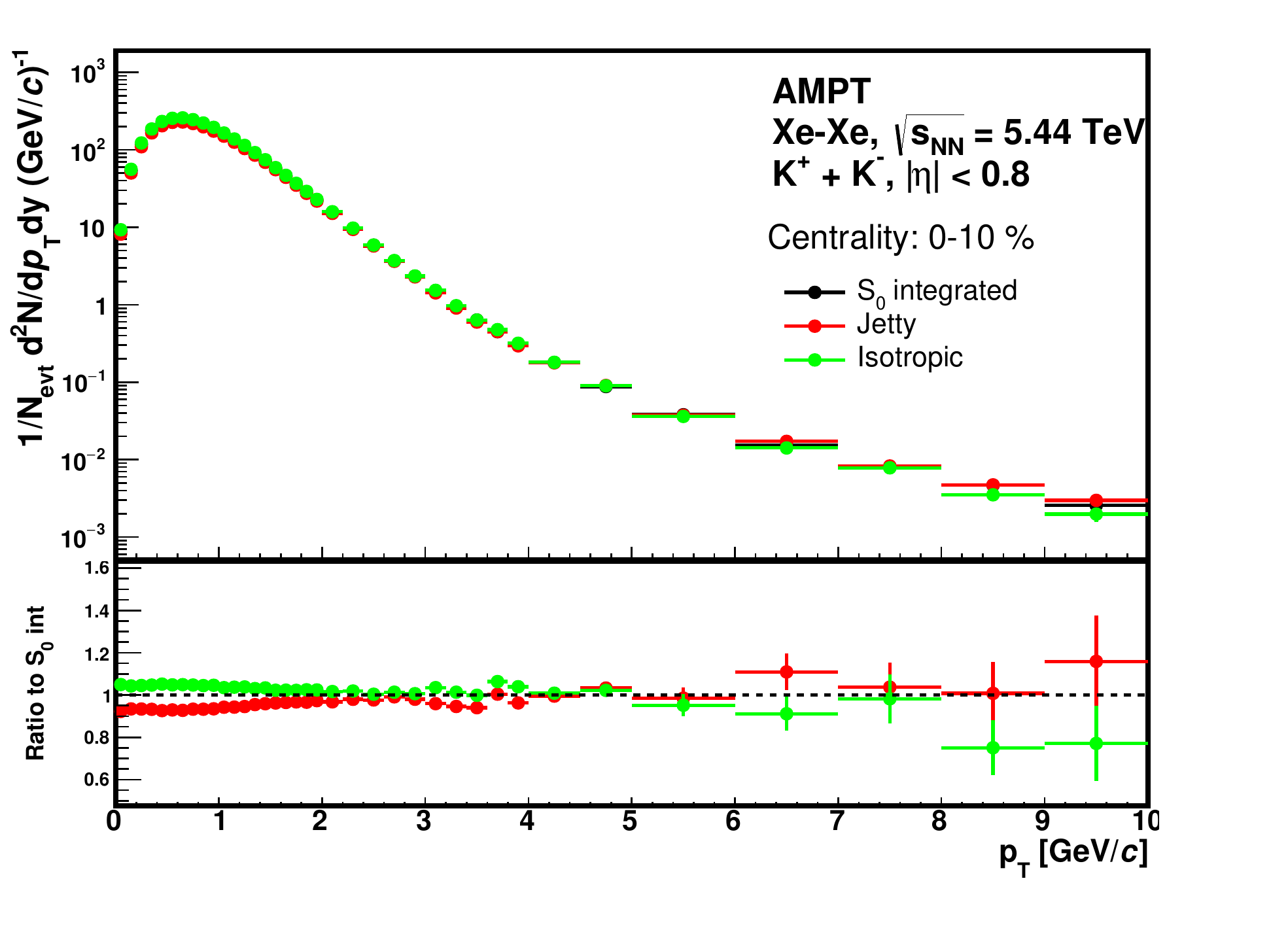}
		\caption{\textbf{AMPT: \kao}}
		\label{Pythia_pion0a}
	\end{subfigure}
	\begin{subfigure}[b]{0.30\linewidth}
	\includegraphics[width=6cm,height=5cm]{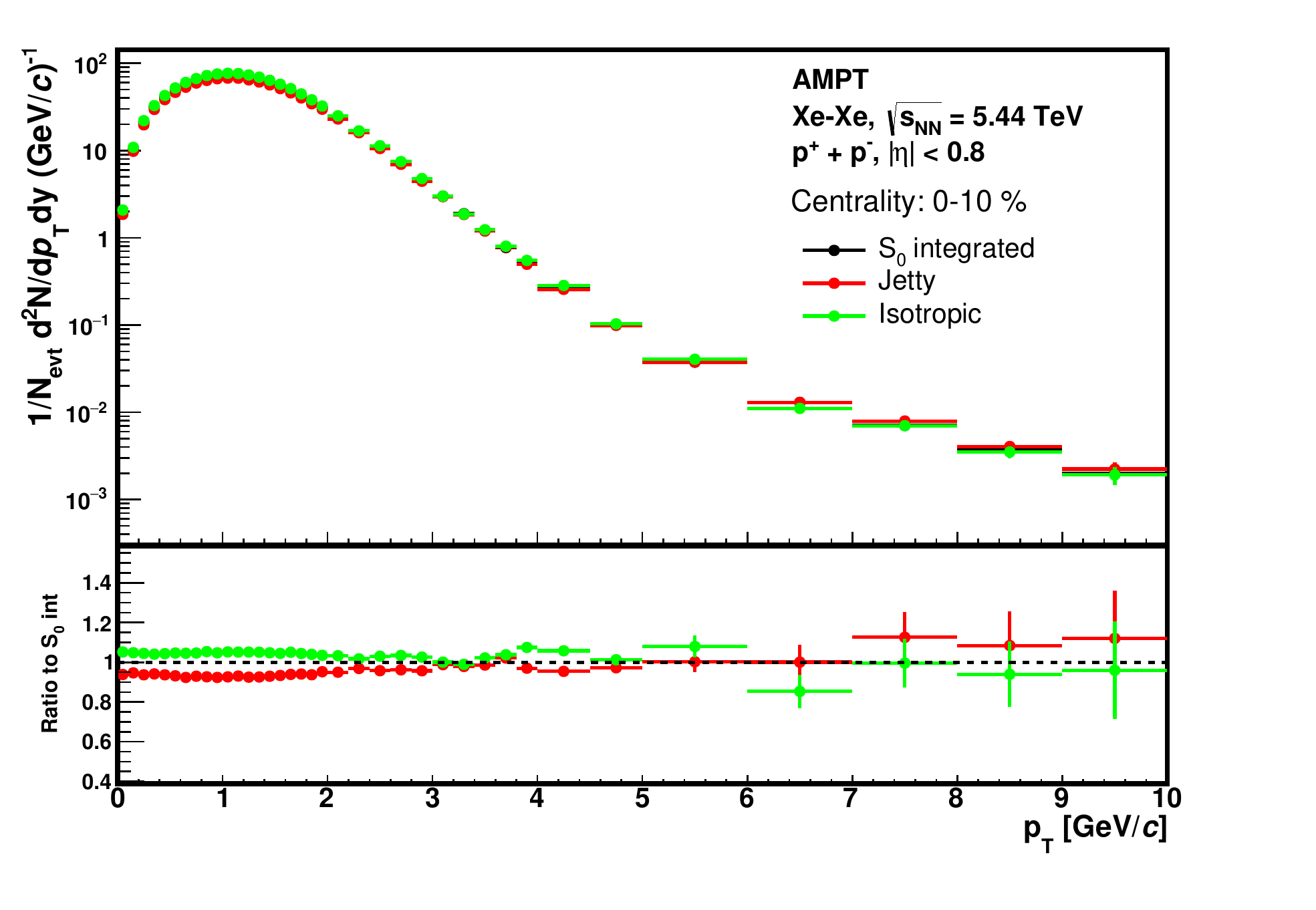}
	\caption{\textbf{AMPT: \pro}}
	\label{AMPT_pion3a}	
\end{subfigure}
\begin{subfigure}[b]{0.30\linewidth}
	\includegraphics[width=6cm,height=5cm]{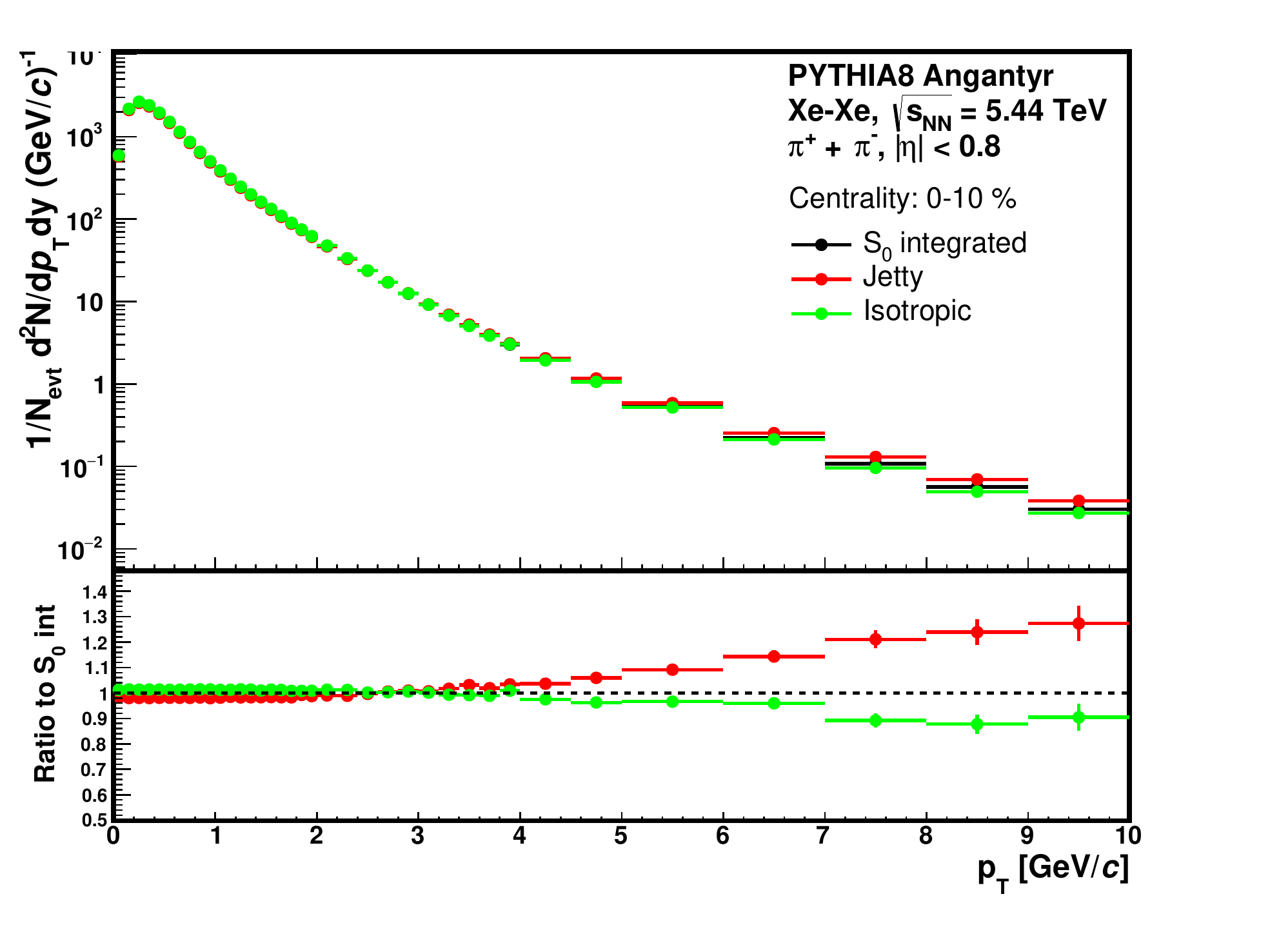}
	\caption{\textbf{PYTHIA8: \pio}}
	\label{Pythia_pion3a}
\end{subfigure}
	\begin{subfigure}[b]{0.30\linewidth}
	\includegraphics[width=6cm,height=5cm]{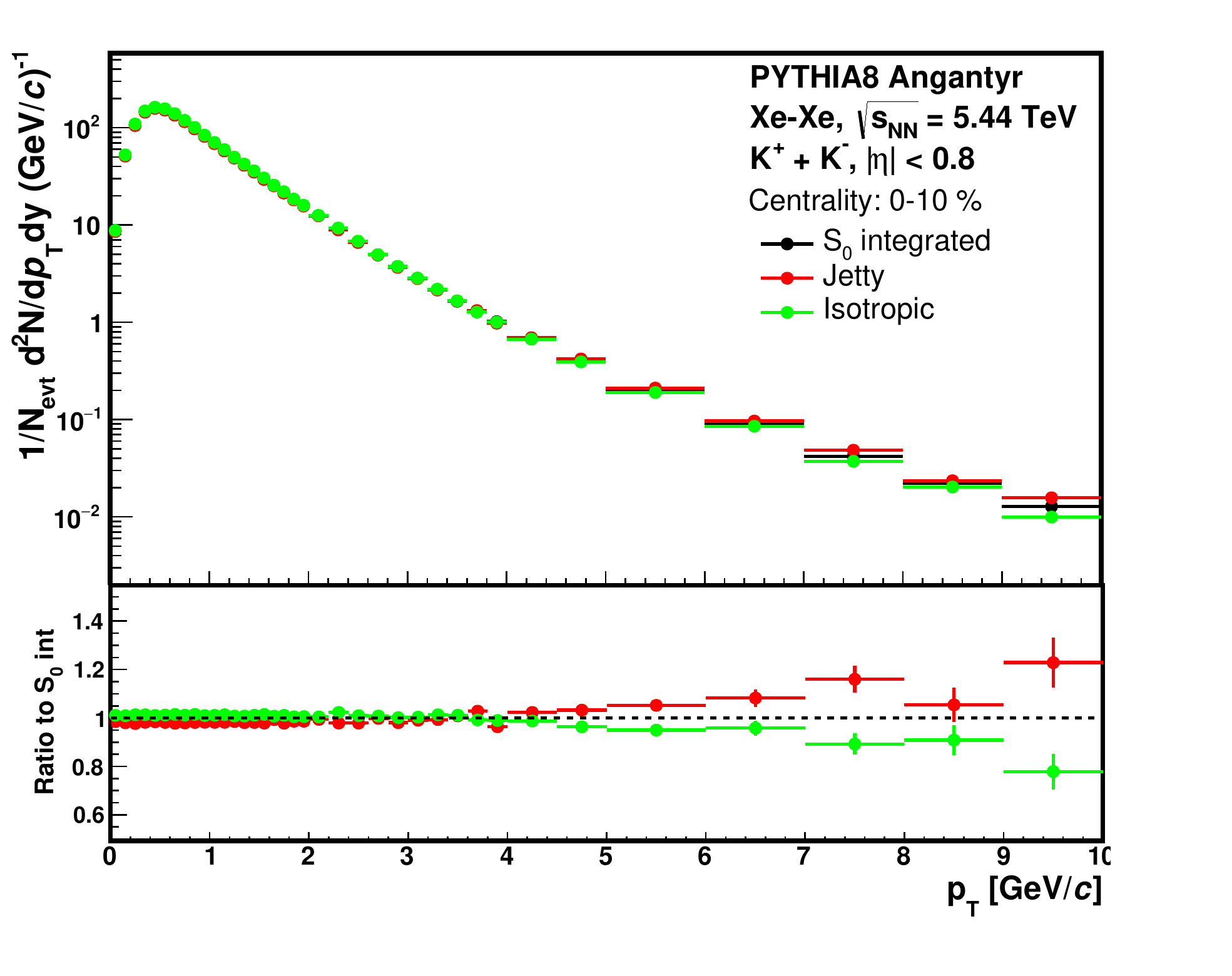}
	\caption{\textbf{PYTHIA8: \kao}}
	\label{AMPT_pion6a}	
\end{subfigure}
\begin{subfigure}[b]{0.30\linewidth}
	\includegraphics[width=6cm,height=5cm]{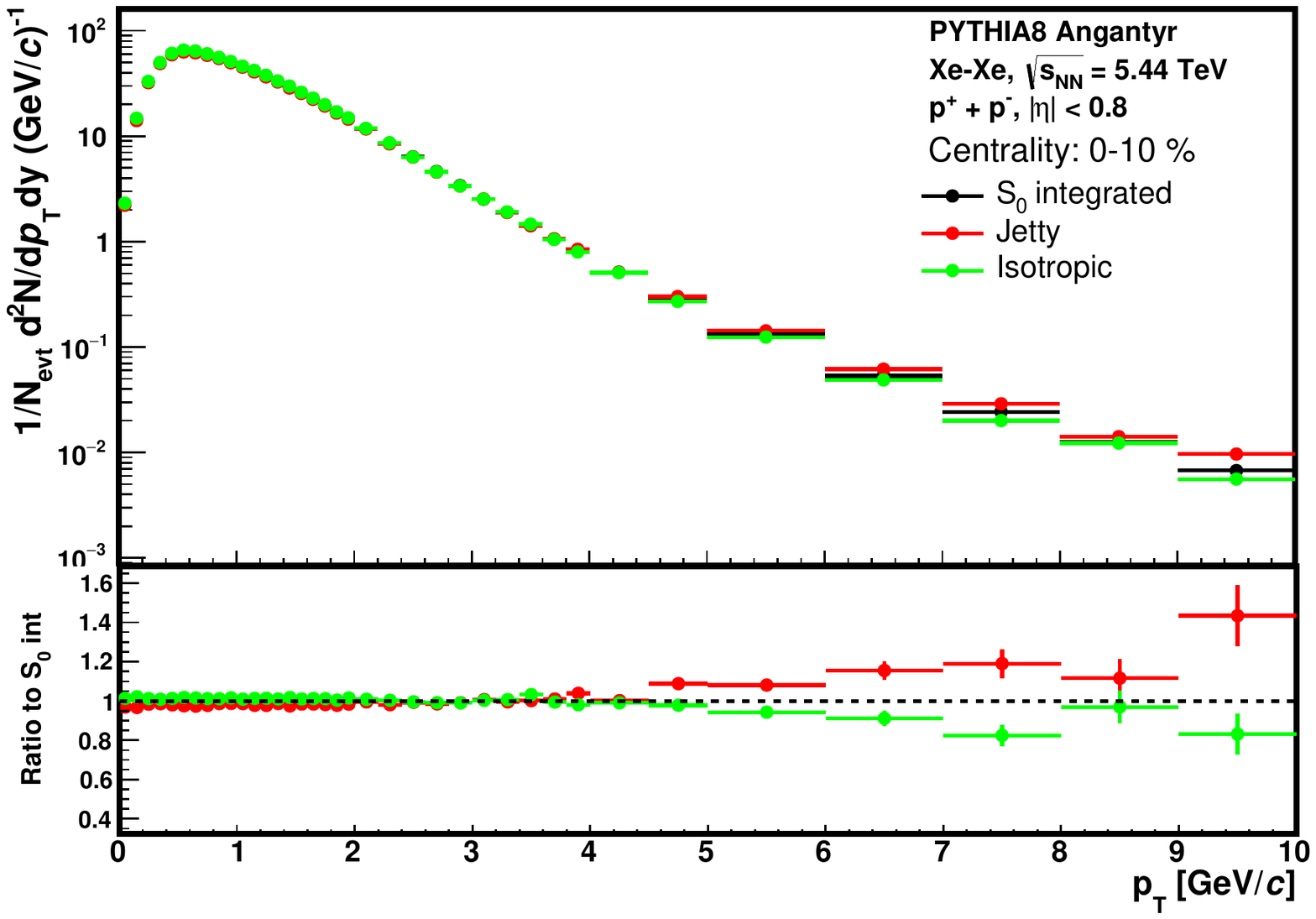}
	\caption{\textbf{PYTHIA8: \pro}}
	\label{Pythia_pion6a}
\end{subfigure}	
\caption{\pts spectra of \pio, \kao and \pro in the $\cent$ bin 0-10$\%$ bin for AMPT (left) PYTHIA8 (right) event generators. The lower panel of the respective figures show the ratio of \pts-spectra for isotropic and jetty events w.r.t \so integrated events. The green markers are for isotropic events, red markers are jetty events, and black represent \so integrated events.}
	\label{Cent_0-10}
\end{figure}

\begin{figure}[h!] 
	\centering
	\begin{subfigure}[b]{0.30\linewidth}
		\includegraphics[width=6cm,height=5cm]{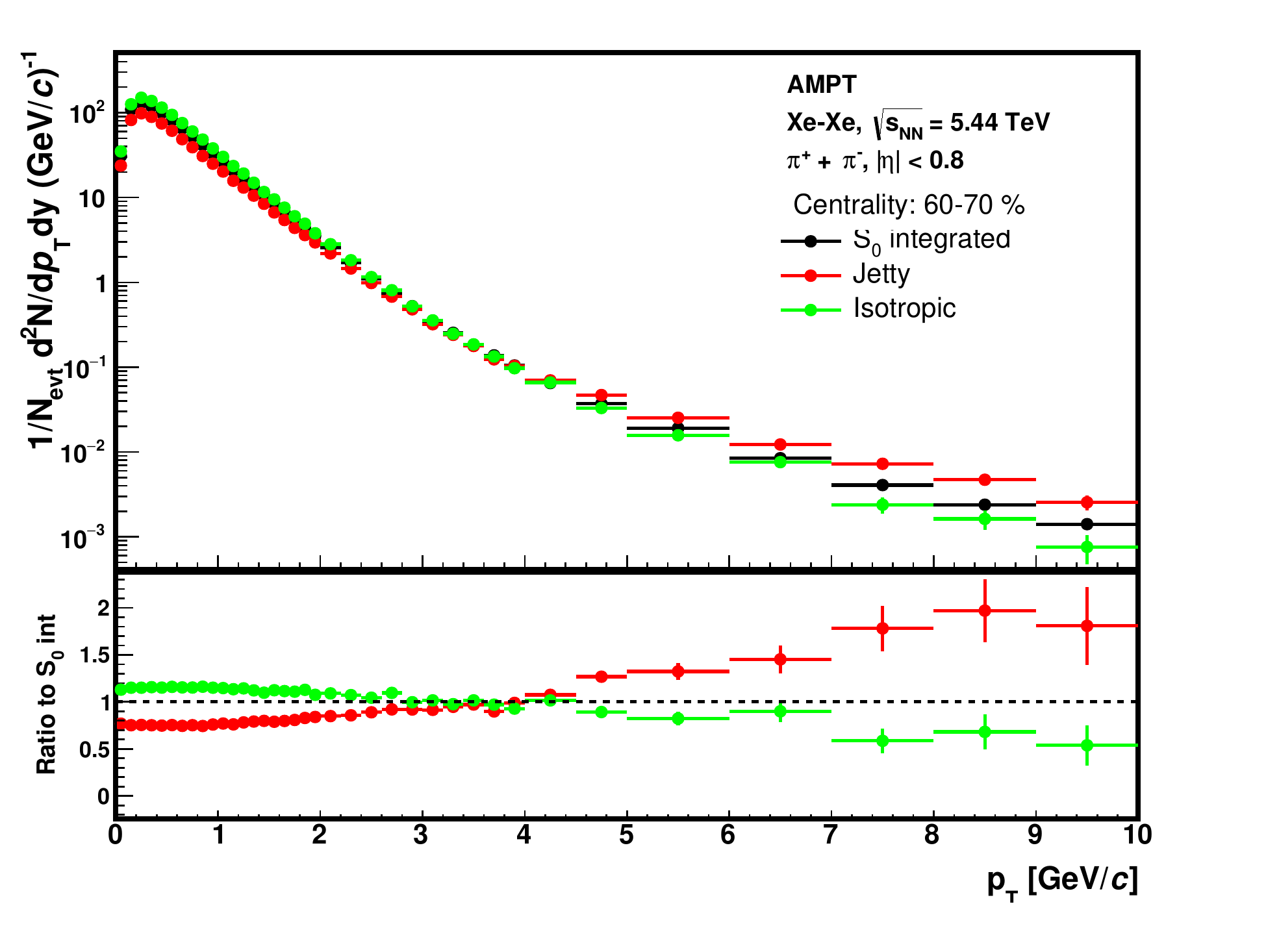}
		\caption{\textbf{AMPT: \pio}}
		\label{AMPT_pion0b}	
	\end{subfigure}
	\begin{subfigure}[b]{0.30\linewidth}
		\includegraphics[width=6cm,height=5cm]{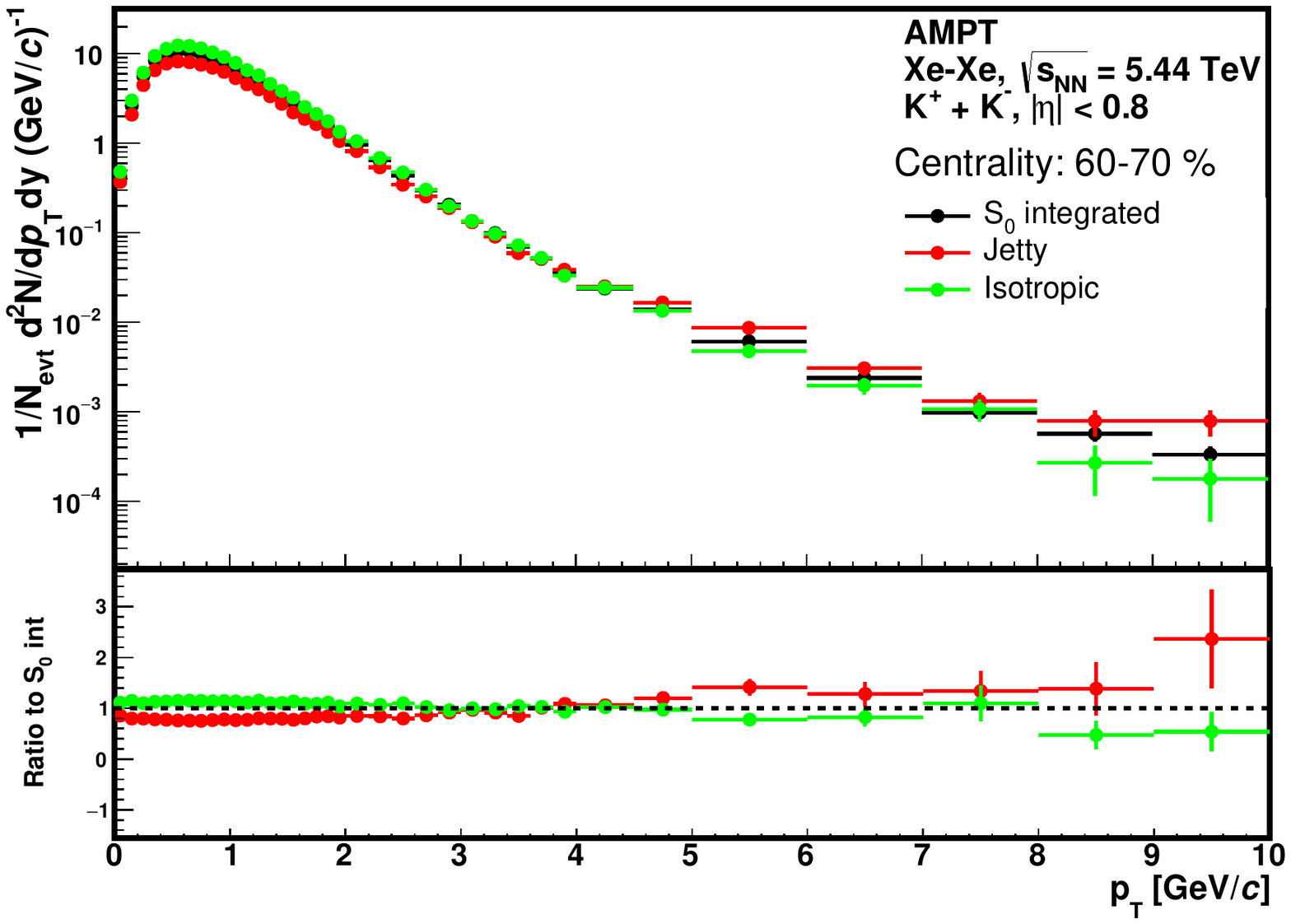}
		\caption{\textbf{AMPT: \kao}}
		\label{Pythia_pion0b}
	\end{subfigure}
	\begin{subfigure}[b]{0.30\linewidth}
		\includegraphics[width=6cm,height=5cm]{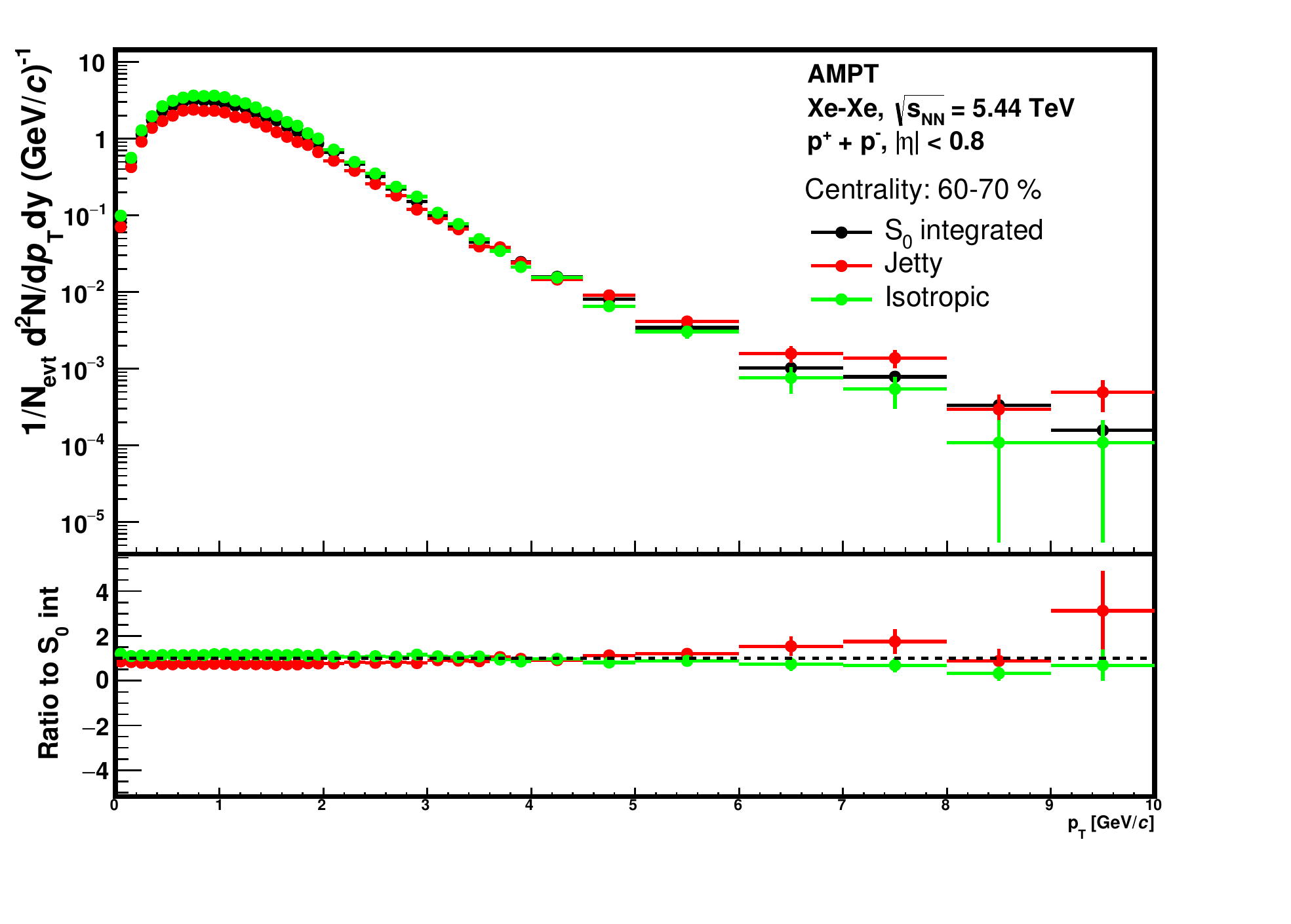}
		\caption{\textbf{AMPT: \pro}}
		\label{AMPT_pion3b}	
	\end{subfigure}
	\begin{subfigure}[b]{0.30\linewidth}
		\includegraphics[width=6cm,height=5cm]{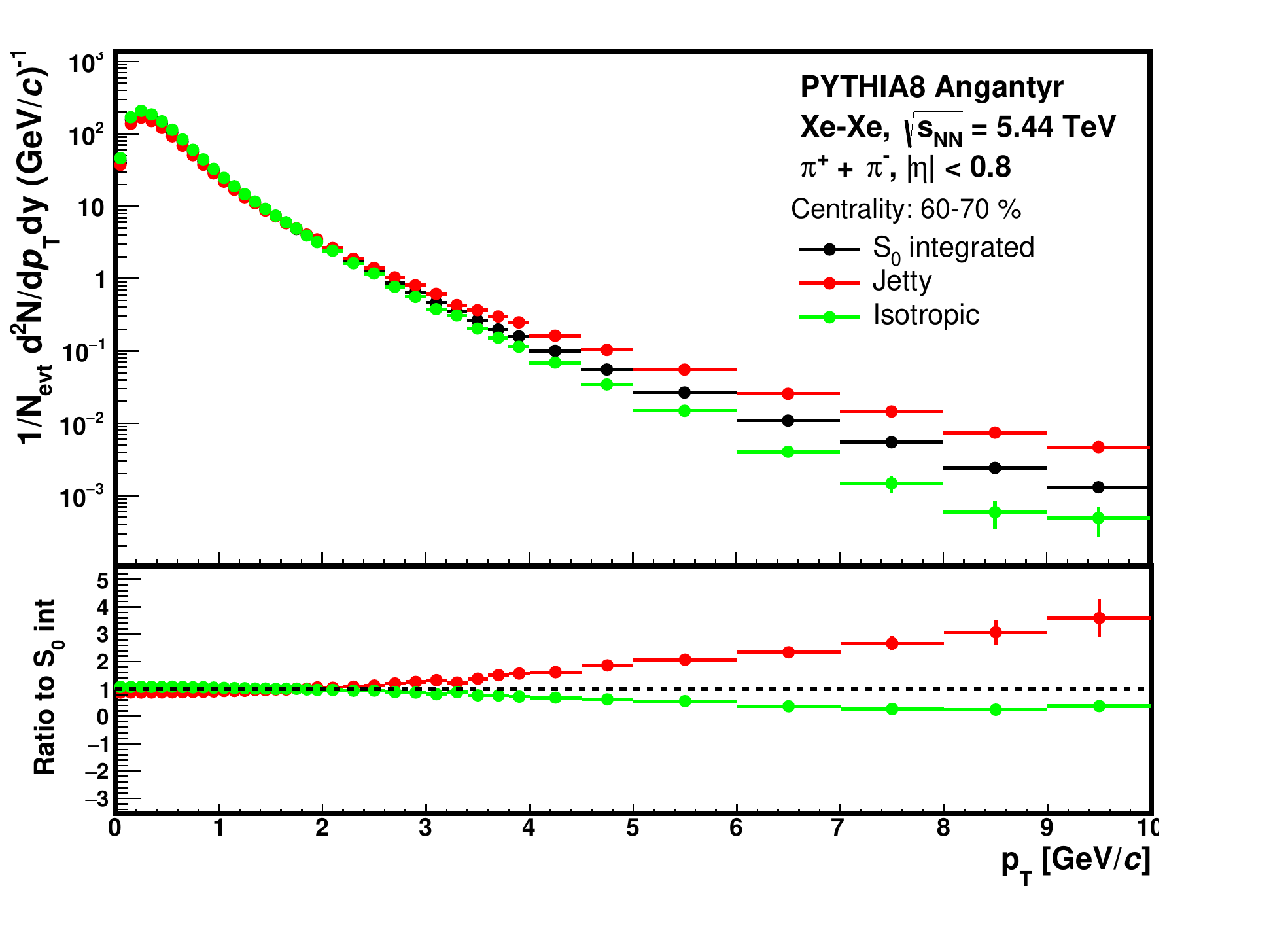}
		\caption{\textbf{PYTHIA8: \pio}}
		\label{Pythia_pion3b}
	\end{subfigure}
	\begin{subfigure}[b]{0.30\linewidth}
		\includegraphics[width=6cm,height=5cm]{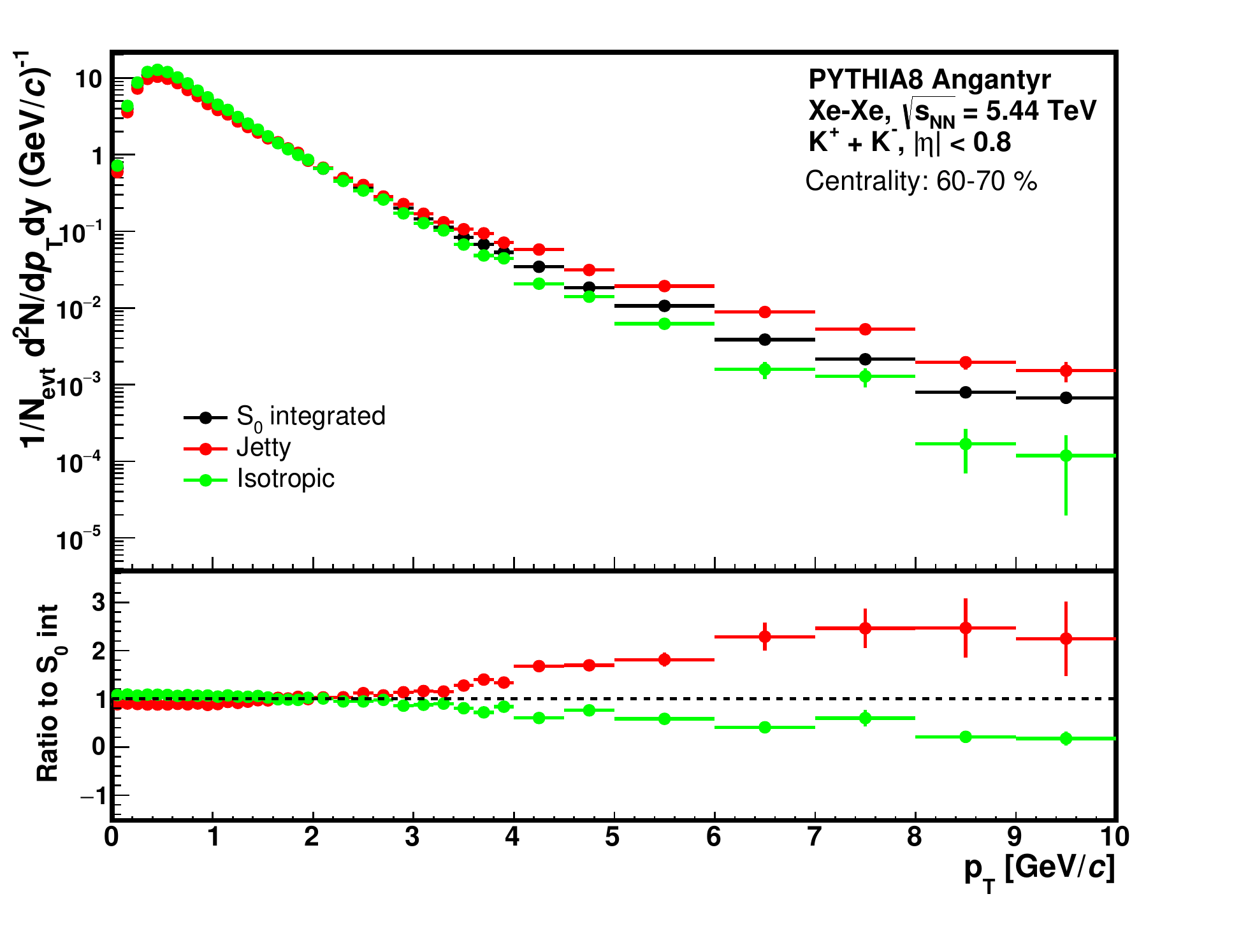}
		\caption{\textbf{PYTHIA8: \kao}}
		\label{AMPT_pion6b}	
	\end{subfigure}
	\begin{subfigure}[b]{0.30\linewidth}
		\includegraphics[width=6cm,height=5cm]{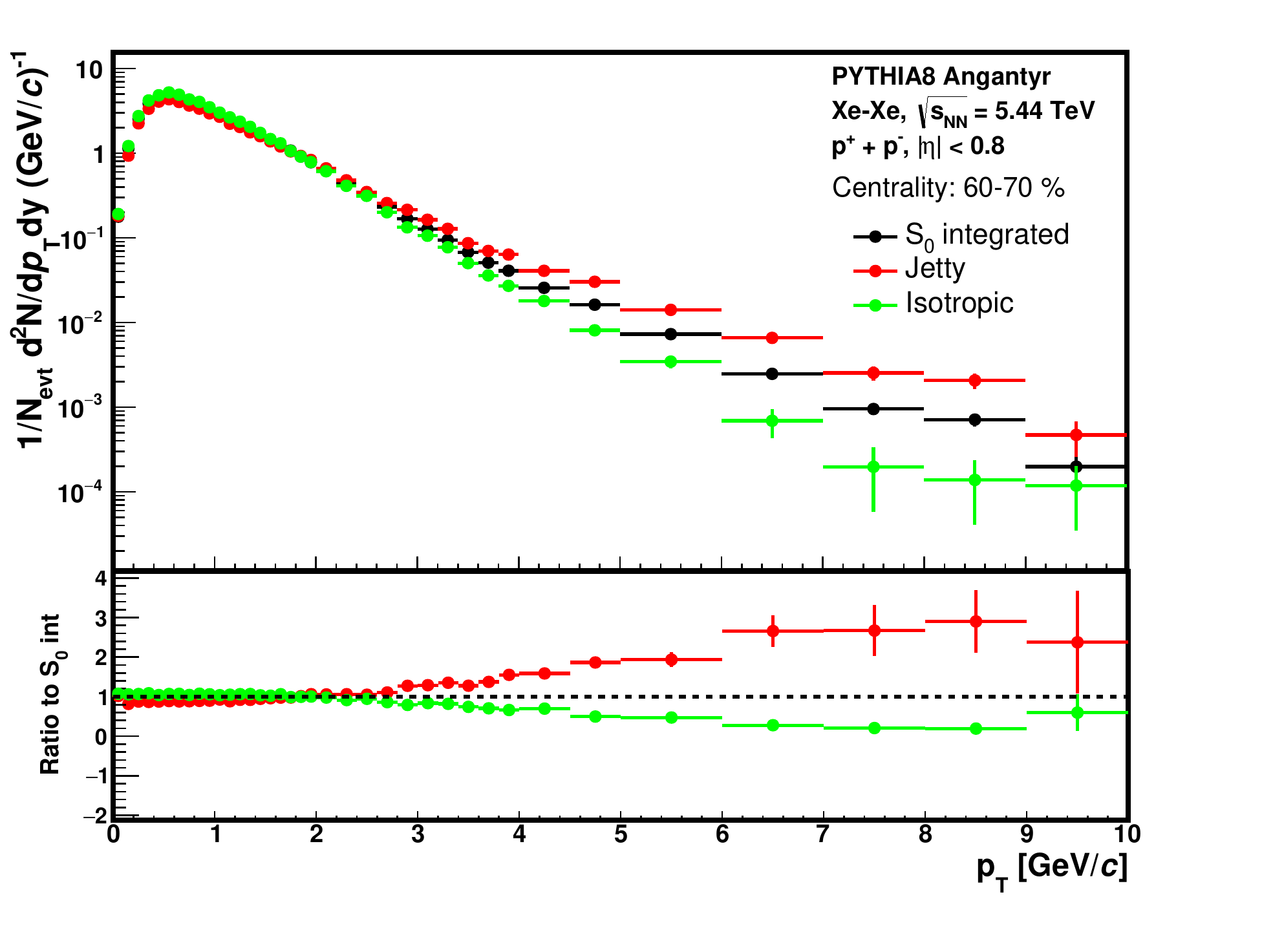}
		\caption{\textbf{PYTHIA8: \pro}}
		\label{Pythia_pion6b}
	\end{subfigure}	
	\caption{\pts spectra of \pio, \kao and \pro in the $\cent$ bin 60-70$\%$ bin for AMPT (left) PYTHIA8 (right) event generators. The lower panel of the respective figures show the ratio of \pts-spectra for isotropic and jetty events w.r.t \so integrated events. The green markers are for isotropic events, red markers are jetty events, and black represent \so integrated events.}
	\label{Cent_60-70}
\end{figure}

\begin{table}[h!]
	\centering
	\resizebox{10cm}{!}{\begin{tabular}{|l|l|l|l|l|}
			\hline
			\multirow{2}{*}{\textbf{Centrality }} &
			\multicolumn{2}{c|}{\textbf{AMPT}} & \multicolumn{2}{c|}{\textbf{PYTHIA8}}\\
			\cline{2-5}
			& jetty & isotropic & jetty& isotropic  \\
			\hline
			0-10\% &0-0.87   & 0.95-1 & 0-0.93 & 0.97-1  \\
			\hline
			10-20\% &0-0.82  & 0.92-1 & 0-0.92 & 0.96-1\\
			\hline
			20-30\% & 0-0.78 & 0.90-1 & 0-0.91 & 0.96-1\\
			\hline
			30-40\% & 0-0.76 & 0.89-1 & 0-0.89 & 0.95-1\\
			\hline
			40-50\% & 0-0.75 & 0.89-1 & 0-0.87 & 0.94-1\\
			\hline
			50-60\% & 0-0.74 & 0.88-1 & 0-0.83 &0.93-1\\
			\hline
			60-70\% & 0-0.72 & 0.87-1 & 0-0.77 & 0.93-1\\
			\hline
		
	\end{tabular}}
	\caption{Table showing the values of \ts for jetty and isotropic ranges for different centrality bins.}
	\label{values}
\end{table}

\begin{figure}[h!]
	\centering
	\includegraphics[scale=0.65]{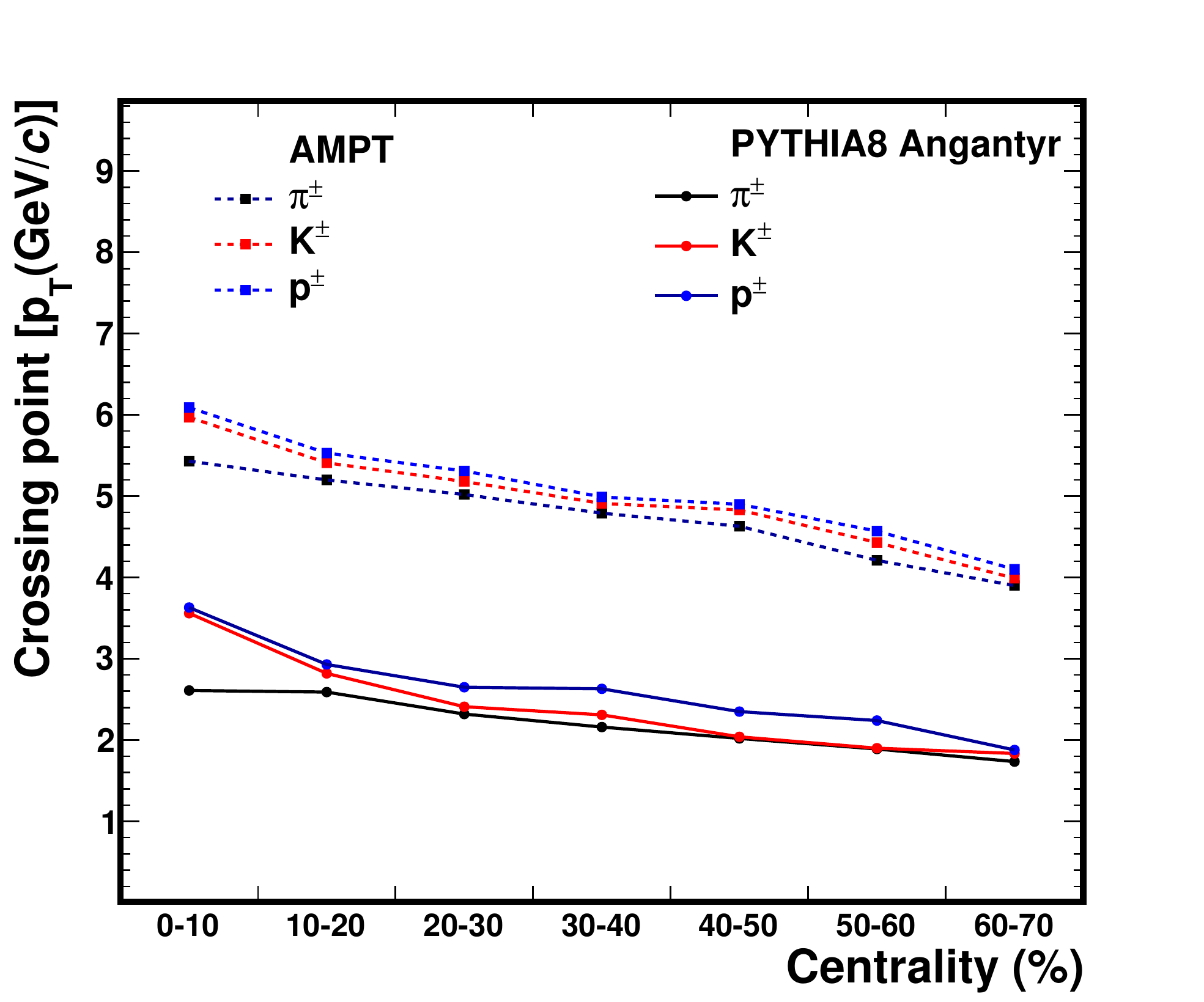}
	\caption{The dependence of ‘crossing point’ on the \cent of the event for PYTHIA8 (filled circles) and AMPT (Square boxes) respectively.}
	\label{crossing}
\end{figure}

The event shape study is done using \ts  in various centrality classes as given in Table \ref{values}. In Figure \ref{Cent_0-10} anf \ref{Cent_60-70}, the \pts spectra of \pio, \kao and \pro for the \cent bin 0-10$\%$ and 60-70$\%$ are shown for various spherocity classes at the mid-rapidity $|\eta| < 0.8$. The different classes of centrality are jetty (which corresponds to lower 20$\%$ of the events in spherocity distribution), isotropic (which corresponds to higher 20$\%$ of the events in spherocity distribution) and integrated(0-1 \ts). The ratio of the \pt-spectra for isotropic and jetty events with respect to the spherocity integrated events ($0 < S_o < 1$) is shown in the lower panels. From the figures, we find that the low \pts regions are dominated by isotropic events rather than jetty events. However, this scenario is reversed when moving to higher \pts. At certain points, called \emph{crossing point}, the jetty event dominates the isotropic event. Therefore, the study of 'crossover points' is of great interest with respect to feasible limits of event-type dominance  and hence related particle generation mechanisms. As we progress from low-multiplicity to high-multiplicities, it has been observed from prior investigations of the light-flavor sector that the 'crossing point' relies on the multiplicity greatly and that the isotropic events are populated over numerous events \cite{Khuntia:2018qox,Bencedi:2018ctm,ALICE:2019bdw}. According to the observations, the QGP-like effects observed in high-multiplicity pp collisions may not be caused by jet-bias effects but instead may be caused by a potential system development, which should be investigated. In the present work, similar studies of the identified particles (\pio, \kao and \pro) in the central pseudorapidity region $|\eta| < 0.8$ in the heavy-ion collision system show that isotropic events are dominant at low centralities (high-multiplicities) and vice-versa. Figure \ref{crossing} shows the the ‘crossing point’ in different \cent ranges for PYTHIA8 and AMPT event generators. The values are also presented in Table \ref{crossing2}. This illustrates how, for two event generators, the contribution of jets to the creation of charged particles varies with centrality conditions. From the Table \ref{crossing2} and Figure \ref{crossing}, a comparison shows the shift of the ‘crossing point’ towards lower-\pts in case of PYTHIA8 Angantyr model indicating the jet dominant contribution in particle production for AMPT model.This means that the contribution of jets is higher in AMPT as compared to the PYTHIA8. The dominance of isotropy in low centrality (high-multiplicity) events and jettiness in high centrality (low-multiplicity) events in charge particle production further suggests a reduction and softening of the jet yields at high charged-particle multiplicity. Therefore in PYTHIA8 Angantyr model, where the high multiplicity events involve the large momentum transfer, the reduction in jet contribution in the particle production may indicate a reduced production of back-to-back jets. However  we should keep in mind that the Angantyr does not include an assumption of a hot thermalised medium, therefore collective effects are absent PYTHIA (Angantyr) whereas AMPT considers collective effects. Also for the Figures \ref{Cent_0-10} and \ref{Cent_60-70}, it is clearly visible that Pythia is clearly able to differentiate the ratio as compared the AMPT where the ratio of jetty and isotropic \pts spectra with respect to the integrated $S_o$ move along the unity in large \pts region. This indicates that the study of the jets production with PYTHIA8 Angantyr and AMPT will form an interesting subject.\\

\begin{table}[h!]
	\centering
	\resizebox{12cm}{!}{\begin{tabular}{|l|l|l|l|l|l|l|}
			\hline
			\multirow{2}{*}{\textbf{Centrality }} &
			\multicolumn{2}{c|}{\textbf{\pio}} & \multicolumn{2}{c|}{\textbf{\kao}}& \multicolumn{2}{c|}{\textbf{\pro}}\\
			\cline{2-7}
			& PYTHIA & AMPT & PYTHIA& AMPT& PYTHIA&AMPT  \\
			\hline
			0-10\% &2.61   & 5.83 & 3.56 & 5.97&3.63&6.09  \\
			\hline
			10-20\% &2.59  & 5.2 & 2.82 & 5.41&2.93&5.53\\
			\hline
			20-30\% &2.32 & 5.02 & 2.41 & 5.18&2.65&5.31\\
			\hline
			30-40\% & 2.16 & 4.79 & 2.31 & 4.91&2.63&4.99\\
			\hline
			40-50\% & 2.02 & 4.63 & 2.04 & 4.83&2.35&4.90\\
			\hline
			50-60\% & 1.89 & 4.21 & 1.90 &4.43&2.24&4.57\\
			\hline
			60-70\% & 1.73 & 3.9 & 1.83 & 3.99&1.88&4.1\\
			\hline
		
	\end{tabular}}
	\caption{Table showing the crossing point (\pts in GeV/c) of jetty and
		isotropic events at mid-rapidity for Xe-Xe at \sqt collisions for PYTHIA8 and AMPT.}
	\label{crossing2}
\end{table}

\section{Summary}
\label{section4}
In conclusion, using A Multi-Phase Transport Model (AMPT) and PYTHIA8 Angantyr, we present the first application of transverse spherocity analysis for Xe-Xe collisions at \sqt. The findings demonstrate that transverse spherocity successfully distinguishes between high-\so and low-\so heavy-ion collision event topologies. From the results, one can see that crossing points occurs at relatively smaller \pts as we go from low \cent to high \cent events. This means that for low \cent events jettiness has a dominance over the isotropiness whereas for high \cent events isotropiness has dominance. Also for PYTHIA8 Angantyr crossing point occurs at much lower \pts as compared to AMPT indicating that jets production occurs at much lower \pts for PYTHIA8 Angantyr as compared to the AMPT. As we know that AMPT considers the formation of medium as well as the collective effects whereas PYTHIA8 Angantyr does not take into account such happenings. Therefore, jet production studies with both models will be an interesting study to do. The results, in our opinion, are extremely positive, and an experimental investigation in this direction would be very beneficial to comprehend the event topology dependence of system dynamics in heavy-ion collisions.

\printnomenclature
\end{document}